\documentstyle[12pt]{article}
\begin{document}
\begin{flushright}
DTP--MSU 95/24 \\ July 28, 1995\\ Hep-th/9507164
\end{flushright}
\hoffset = -1truecm
\voffset = -2truecm
\vskip4cm\begin{center}
{\LARGE\bf
Matrix Dilaton--Axion for Heterotic String in Three Dimensions
 }\\ \vskip1cm
{\bf
D.V. Gal'tsov}
and
{\bf  O.V.  Kechkin}\\
\normalsize  Department of Theoretical Physics,  Moscow State University,\\
\normalsize Moscow 119899, {\bf Russia}\\
\normalsize \date{27 July 1995} \vskip2cm
{\bf Abstract}\end{center}
\begin{quote}
New and surprisingly simple representation is found for the heterotic
string bosonic effective action in three dimensions in terms of
complex potentials. The system is presented as a K\"ahler
$\sigma$--model using complex symmetric $2\times 2$ matrix
(matrix dilaton--axion) which depends linearly  on three Ernst--type
potentials and transforms under $U$--duality via matrix valued $SL(2,R)$.
Two discrete automorphisms relating ten isometries of the target space
(U--duality transformations) are found and used to generate the non--trivial
Ehlers--Harrison sector by a map from the trivial gauge sector. Finite
transformations are obtained in a simple form in terms of complex potentials.
New solution generating technique is used to construct EMDA double--Kerr
solution.

\vskip5mm
\noindent
PASC number(s): 97.60.Lf, 04.60.+n, 11.17.+y
\end{quote}
\newpage

It was found recently that the  low--energy effective theory of the
toroidally compactified heterotic string possesses large internal
symmetries when reduced to three \cite{sht} and two dimensions \cite{bmss}.
This phenomenon was firstly described \cite{gk} within a framework
of the dilaton--axion gravity theory, which is initially
formulated in four dimensions and incorporates a dilaton,
an axion and one $U(1)$ vector field. Two related features are
essential: i) unification of T and S dualities within a larger
(U--duality) group in three dimensions \cite{gk}, ii) symmetric
space nature of the target space of a corresponding
three--dimensional $\sigma$--model \cite{g}, which
renders further two--dimensional reduction of the theory
to be fully integrable \cite{maz, bgm}. It is worth noting that derivation
of these properties includes dualization of certain variables,
so that symmetry group acts on potential space \cite{nk} rather
than on the primary metric and vector field variables. U--duality group
for dilaton--axion gravity (Einstein--Maxwell--dilaton--axion
(EMDA) coupled system) in three dimensions is $SO(2,3)$, while the
target space is the coset $SO(2,3)/(SO(3)\times SO(2))$ \cite{g}.
The corresponding symmetry of the full ten--dimensional toroidally
compactified heterotic string effective theory is $SO(8,24)$ \cite{sht},
the EMDA truncation arising when only one vector and no moduli fields
excited.

In view of isomorphism $SO(2,3)\sim Sp(4,R)$ the EMDA system in three
dimensions admits a symplectic formulation \cite{gkt} which results in the
introduction of the ``matrix axion--dilaton'' combining
six real potential space variables into one complex symmetric
$2\times 2$ matrix. Under action of U--duality
this object undergoes matrix valued $SL(2,R)$ transformations.
In terms of complex variables three--dimensional
formulation is equivalent to a K\"ahler $\sigma$--model.
Three complex variables introduced in \cite{gk, gkt} may
be regarded as the EMDA analog of Ernst potentials
in the Einstein--Maxwell (EM) theory \cite{er}.
Due to $Sp(4,R)$ symmetry, various forms of the matrix
dilaton--axion in terms of these variables are possible.
The purpose of this paper is to present a new simple expression
for matrix dilaton--axion which is linear in the Ernst--type variables. This
representation reveals useful discrete symmetries and gives a concise
and unified form of U--duality transformations.

We deal with the EMDA four--dimensional action
\begin{equation}
S=\frac{1}{16\pi}\int \left\{-R+2\left|\partial z
(z-{\bar z})^{-1}\right|^2 +
i\left(z{\cal F}^2+c.c\right)\right\}
\sqrt{-g}d^4x,
\end{equation}
where ${\cal F}=(F+i{\tilde F})/2,\;
{\tilde F}^{\mu\nu}=\frac{1}{2}E^{\mu\nu\lambda\tau}F_{\lambda\tau}$,
and
\begin{equation}
z=\kappa+ie^{-2\phi}
\end{equation}
is the complex dilaton--axion field. In physical terms the EMDA action
may be regarded as Brans--Dicke--type generalization of the EM theory
satisfying additional requirement of continuous electric--magnetis duality.
In the purely dilatonic generalization the Abelian electric--magnetic
duality of the EM theory is broken, while an axion helps to restore
this symmetry extending it to non--Abelian $S$--duality \cite{sd},
$SL(2,R)$. As a result eight--parametric symmetry $SU(2,1)$ of the EM
system \cite{ki} is transformed into ten--parametric group $Sp(4,R)$
in the EMDA case.

To reduce the action (1) to three dimensions consider a space--time
admitting a non--null (time--like for definitness) Killing symmetry,
so that in adapted coordinates
\begin{equation}
ds^2=g_{\mu\nu}dx^\mu dx^\nu=f(dt-\omega_idx^i)^2-\frac{1}{f}h_{ij}dx^idx^j.
\end{equation}
Here metric of the three--space $h_{ij},\;(i, j=1, 2, 3)$, a rotation
one--form $\omega_i $ and a three--dimensional conformal factor
$f$ depend on the 3--space coordinates $x^i$. Maxwell tensor is parametrized
by a time component $v$ of the vector potential
\begin{equation}
F_{i0}=\frac{1}{\sqrt{2}}\partial_iv,
\end{equation}
and by a magnetic potential $u$
\begin{equation}
e^{-2\phi}F^{ij}+\kappa {\tilde F}^{ij}=\frac{f}{\sqrt{2h}}\epsilon^{ijk}
\partial_ku.
\end{equation}
Then a twist potential $\chi$ is introduced to generate $\omega_i$ in
conformity with the Einstein constraint
\begin{equation}
\nabla \chi=u\nabla v -v\nabla u -f^2\nabla \times \omega,
\end{equation}
where $\nabla$ is the three--dimensional covariant derivative.

 From six real variables entering a three--dimensional problem as a set
of scalar fields one can build, apart from (2), two more complex potentials
\begin{eqnarray}
\Phi&=&u-zv, \nonumber \\
E&=&if-\chi+v\Phi,
\end{eqnarray}
in terms of which after reduction to three dimensions we get
the K\"ahler $\sigma$--model with the target space metric
\begin{equation}
dl^2=\frac{1}{2f^2}\left|dE+\frac{2{\rm Im}\Phi}{{\rm Im}z} d\Phi-
\left(\frac{{\rm Im}\Phi}{{\rm Im}z}\right)^2dz\right|^2-
\frac{1}{f{\rm Im}z}\left|d\Phi-\frac{{\rm Im}\Phi}{{\rm Im}z}dz \right|^2+
\frac{|dz|^2}{2({\rm Im}z)^2}.
\end{equation}
This formula can be obtained by substituting real potentials in terms
of complex ones into the line element found in \cite{gk}.

The representation of this target space as  coset $Sp(4,R)/U(2)$ reads
\begin{equation}
{dl}^2=\frac{1}{4}Tr(\omega^2),\quad
\omega=dMM^{-1},
\end{equation}
where
\begin{equation}
M=\left(\begin{array}{crc}
P^{-1}&P^{-1}Q\\
QP^{-1}&P+QP^{-1}Q\\
\end{array}\right),
\end{equation}
$P,\quad Q$ being  real symmetric $2\times 2$ matrices.
Their complex combination
\begin{equation}
Z=Q+iP,
\end{equation}
can be called ``matrix dilaton--axion''  for the following reason.
Under an action of  $G\in Sp(4, R)$ on the coset
\begin{equation}
M\rightarrow G^TMG
\end{equation}
matrix $Z$ transforms as
\begin{equation}
Z \rightarrow Z+R,
\end{equation}
\begin{equation}
Z \rightarrow S^TZS,
\end{equation}
\begin{equation}
Z^{-1} \rightarrow Z^{-1}+L,
\end{equation}
where $R,\; S,\;L$ are real $2\times 2$ matrices comprising ten real
parameters of the isometry group $Sp(4,R)$ of (8) (here $R^T=R,\; L^T=L,\;
S$ arbitrary non--degenerate). This may be
interpreted as matrix--valued $SL(2, R)$ transformations: usual $SL(2,R)$
transformations act on scalar dilaton--axion $z$ (2) under S--duality
subgroup in a similar way with $R,\; S,\;L$ real numbers.
Three--dimensional action can now be rewritten in terms of the matrix
dilaton--axion as follows
\begin{equation}
S_\sigma=\int\left\{-{\cal R}+2{\rm Tr}(J{\bar J}\right\},\quad
J=\nabla Z (Z-{\bar Z})^{-1},
\end{equation}
where ${\cal R}$ is the Ricci scalar of a three space.
Comparing this with the initial four--dimensional action (1) we see
that now three pairs of real variables are conveniently incorporated
into a matrix which enters the action (16) in the same way as scalar
dilaton--axion $z$ enters (1).

Due to an invariance of the action under (13)--(15), a concrete form
of matrix dilaton--axion in terms of complex potentials is not unique.
In \cite{gkt} a particular form was obtained using the correspondence
between target space Killing vectors written in real and $Z$ variables.
All elements of the matrix $Z$ were expressed as ratios of two
second order polinomials of complex potentials.
It turns out that using a sequence of $Sp(4,R)$ transformations one can
simplify this form drastically, and ultimately to find $Z$ as a linear
function of $E,\; \Phi,\;z$. This involves the following chain of operations.
First, one makes a shift $\Phi \rightarrow \Phi -1$, which is a (magnetic)
gauge transformation. Then the full matrix $Z$ is shifted on a real matrix
$Z \rightarrow Z +\sigma_z$ (Pauli matrix), this obviously leaves
invariant the action (16). This is followed by an inversion
$Z \rightarrow Z^{-1}$, (also symmetry of (16)). Finally suitable
$S$--transformation is made followed by change of a sign of $Z$.
This results in an extremely simple final form of $Z$ (here we use ``Latin''
Ernst potential related to one introduced in \cite{gkt} as $E=i{\cal E}$)
\begin{equation}
Z=\left(\begin{array}{crc}
E&\Phi\\
\Phi&-z\\
\end{array}\right).
\end{equation}

The action (16) is manifestly invariant under $Sp(4,R)$ transformations
written in the ``matrix $SL(2,R)$ form'' (13)--(15). In particular,
one useful discrete $S$--transformation is
\begin{equation}
Z\rightarrow Z^{'}=\varepsilon Z \varepsilon,\quad
\varepsilon=\left(\begin{array}{crc}
0&1\\
-1&0\\
\end{array}\right),
\end{equation}
(``prime'' operation). In terms of complex potentials it reads
\begin{equation}
z^{'}=E,\quad  E^{'}=z,\quad \Phi^{'}=\Phi.
\end{equation}
Anoter useful discrete symmetry operation, which will be called
``double priming'', is the inversion
\begin{equation}
Z \rightarrow Z^{''}=Z^{-1}.
\end{equation}
It corresponds to
\begin{equation}
E^{''} = \frac{z}{Ez+\Phi^2},\quad
z^{''} = \frac{E}{Ez+\Phi^2},\quad
\Phi^{''} =\frac{\Phi}{Ez+\Phi^2}.
\end{equation}
These two discrete maps open a convenient way to express all non--trivial
isometries of the K\"ahler space (8) through three gauge and one scale
transformations which are easy to find (see \cite{gk}).
Let us do this firstly in the infinitesimal form. The set of three
gauge (electric $e$, magnetic $m$, and gravitational $g$) Killing
vectors in terms of complex potentials read
\begin{equation}
K_e=2\Phi\partial_E - z\partial_\Phi +c.c.,
\end{equation}
\begin{equation}
K_m=\partial_\Phi +c.c.,
\end{equation}
\begin{equation}
K_g=\partial_E +c.c.,
\end{equation}
and the scale generator is
\begin{equation}
K_s=2E\partial_E+ \Phi\partial_\Phi +c.c..
\end{equation}

Ehlers--Harrison--type generators \cite{eh, ki} were  found for EMDA theory
in \cite{gk} using real variables, in which they are rather involved.
Now they can be obtained simply by ``priming'' gauge generators
\begin{equation}
K_{H_1}= K_e^{'}=2\Phi\partial_z - E\partial_\Phi +c.c.,
\end{equation}
\begin{equation}
K_{H_2}= K^{''}_m=(Ez-\Phi^2)\partial_\Phi- 2\Phi(z\partial_z +
E\partial_E) +c.c.,
\end{equation}
\begin{equation}
K_{E}= K^{''}_g=\Phi^2\partial_z-E(E\partial_E+\Phi\partial_\Phi) +c.c..
\end{equation}

Two generators of $S$--duality subgroup corresponds to the primed $g$ and
$s$ transformations
\begin{equation}
K_{d_1}= K_g^{'}=\partial_z +c.c,
\end{equation}
\begin{equation}
K_{d_3}= K_s^{'}=2z\partial_z +\Phi\partial_\Phi + c.c,
\end{equation}
while the remaining one may be obtained by priming the Ehlers generator
\begin{equation}
K_{d_2}= K^{'}_E=\Phi^2\partial_E-z(z\partial_z+\Phi\partial_\Phi) +c.c..
\end{equation}
This completes the list of ten $sp(4,R)\;U$--duality generators of the
three--dimensional reduction of the EMDA theory.

The same procedure can now be applied in order to obtain {\em finite}
$U$--duality elements in terms of complex potentials.
Starting with gravitational
\begin{equation}
E=E_0+\lambda,\quad \Phi=\Phi_0,\quad z=z_0,
\end{equation}
electric
\begin{equation}
E=E_0-2\lambda \Phi_0-\lambda^2 z_0,\quad
\Phi=\Phi_0+\lambda z_0,\quad z=z_0,
\end{equation}
and magnetic
\begin{equation}
E=E_0,\quad \Phi=\Phi_0+\lambda,\quad z=z_0,
\end{equation}
gauge and a scale transformation
\begin{equation}
E=e^{2\lambda}E_0,\quad \Phi=e^{\lambda}\Phi_0,\quad z=z_0,
\end{equation}
which can be easily found by integrating infinitesimal transformations
(22)--(25), one can find the remaining six group elements using priming
operations (19), (21). The $S$--duality subgroup will read
\begin{eqnarray}
&& d_1:\qquad E=E_0,\quad \Phi=\Phi_0,\quad z=z_0+\lambda,\nonumber\\
&& d_2:\qquad E=E_0+\lambda\frac{\Phi_0^2}{1+\lambda z_0},\quad
\Phi=\frac{\Phi_0}{1+\lambda z_0},\quad
z=\frac{z_0}{1+\lambda z_0},\\
&& d_3:\qquad E=E_0,\quad \Phi=e^{\lambda}\Phi_0,\quad
z=e^{2\lambda}z_0,\nonumber
\end{eqnarray}
the electric and magnetic Harrison transformations are
\begin{equation}
E=E_0,\quad \Phi=\Phi_0+\lambda E_0,\quad
z=z_0-2\lambda \Phi_0-\lambda^2 E_0,
\end{equation}
\begin{equation}
E=\frac{E_0}{(1+\lambda\Phi_0)^2+\lambda^2 E_0 z_0},\quad
\Phi=\frac{\Phi_0(1+\lambda\Phi_0)+\lambda E_0 z_0}
{(1+\lambda\Phi_0)^2+\lambda^2 E_0 z_0},\quad
z=\frac{z_0}{(1+\lambda\Phi_0)^2+\lambda^2 E_0 z_0}.
\end{equation}
while the Ehlers--type transformation reads
\begin{equation}
E=\frac{E_0}{1+\lambda E_0},\quad
\Phi=\frac{\Phi_0}{1+\lambda E_0},\quad
z=z_0+\frac{\lambda\Phi_0^2}{1+\lambda E_0}.
\end{equation}
It can be observed, that the most involved magnetic Harrison--type
transformation (38) becomes really complicated if one attempts to
separate real and imaginary parts (the result is given in \cite{gk}).
An apparent asymmetry between electric and magnetic Harrison transformations
(contrary to the case of purely EM system \cite{eh, ki})
is due to the non--Abelian
nature of the electric--magnetis duality subgroup (36). However in
the present form all finite transformations are quite easy to apply,
additional simplification may be achieved by use of the prime operations
described above.

But even greater simplification comes from the direct use
of the matrix dilaton--axion $Z$. It can be checked that ten transformations
listed above correspond to ``matrix--valued $SL(2,R)$'' transformations
(13)--(15) with the following matrix parameters
\begin{equation}
R=
\left(\begin{array}{crc}
\lambda_g&\lambda_m\\
\lambda_m&-\lambda_{d_1} \\
\end{array}\right),\quad
S=
\left(\begin{array}{crc}
e^{\lambda_s}&\lambda_{H_1}\\
-\lambda_e&e^{\lambda_{d_3}} \\
\end{array}\right),\quad
L=
\left(\begin{array}{crc}
\lambda_E&\lambda_{H_2}\\
\lambda_{H_2}&-\lambda_{d_2} \\
\end{array}\right).
\end{equation}
Here all real group parameters are equiped with indices corresponding
to the above listed transformations.

Using the rules (13)--(15) one
can show that $R,\;L$ and $S$ transformations form subgroups of the
full symmetry group. Moreover, $R$ and $S$ together form a subgroup,
as well as $L$ and $S$ (but not $R$ and $L$). Together with symmetry
under priming (18) this gives rise to a simple generating method.
Let $Z_0$ correspond to an asymptotically flat solution,
$Z_0(\infty)=Z_{0\infty}$. Then under action of the ($S, L$) subgroup
\begin{equation}
Z-Z_{\infty}=S^T(Z_0-Z_{0\infty})S,\quad {\rm Im}Z_{\infty}
=S^T{\rm Im}Z_{0\infty}S.
\end{equation}
Fixing gauges of both seed and resulting solutions so that
\begin{equation}
Z_{\infty}=
\left(\begin{array}{crc}
i&0\\
0&-z_{\infty}\\
\end{array}\right),
\quad
Z_{0\infty}=
\left(\begin{array}{crc}
i&0\\
0&-i\\
\end{array}\right),
\end{equation}
one finds for $S$ the following parametrization
\begin{equation}
S=
\left(\begin{array}{crc}
\cosh \theta &e^{-\phi_{\infty}}\sinh \theta\\
\sinh \theta &e^{-\phi_{\infty}}\cosh \theta\\
\end{array}\right).
\end{equation}

Consider an axisymmetric three-space
\begin{equation}
dl_3^2=h_{ij}dx^idx^j=e^{2\Gamma}(d\rho^2+d\zeta^2)+
\rho^2d\varphi^2.
\end{equation}
If some solution is uncharged $\Phi=0$, then both $E$ and $z$ satisfy
the same vacuum Ernst equation (cf.(19)).
In this case $\Gamma=\gamma+\gamma^{'}$ where $\gamma$ satisfies
the pair of equations
\begin{equation}
\gamma_{,\rho}=\frac{\rho}{(2{\rm Im}E)^2}\left(E_{,\rho}{\bar E}_{,\rho}-
E_{,\zeta}{\bar E}_{,\zeta}\right),\quad
\gamma_{,\zeta}=\frac{\rho}{2({\rm Im}E)^2}E_{,(\rho}{\bar E}_{,\zeta)},
\end{equation}
while $\gamma^{'}$ satisfies similar primed equations (i.e. with
$E\rightarrow z$). Now if one takes a seed solution with $\Phi=0$,
specified by an Ernst potential $E_0$, one can generate another solution
with $\Phi=0$ adding similar $z_0$ which is also a solution to the vacuum
Ernst equation. Let us specify
\begin{equation}
E_0=i\left(1-\frac{2\mu_1}{R_1}\right),\quad
z_0=i\left(1-\frac{2\mu_2}{R_2}\right),
\end{equation}
where
\begin{equation}
R_b=r_b+ia_b\cos\theta_b, \;\;b=1, 2,
\end{equation}
and $r_b,\; \theta_b$ are related to Weyl coordinates as
\begin{equation}
\rho=\left((r_b-\mu_b)^2+\sigma_b^2\right)^{1/2}\sin\theta_b, \quad
\zeta=\zeta_b+(r_b-\mu_b)\cos\theta_b,
\end{equation}
where $\sigma_b^2=a_b^2-\mu_b^2$. Possibility to take $E$ and $z$ centered
at different points follows from the fact that for $\Phi=0$ gravitational
and dilaton--axion sectors decouple. The corresponding metric function
gamma will be $\Gamma=\gamma_1+\gamma_2$ where $\gamma_2=\gamma^{'}$.
Equations (45) then gives
\begin{equation}
e^{\gamma_b}=\frac{(r_b-\mu_b)^2+\sigma_b^2-a_b^2\sin^2\theta_b}
{(r_b-\mu_b)^2+\sigma_b^2\cos^2\theta_b},
\end{equation}
this quantity remains unchanged by the transformation. Hence the transformed
solution will be fully described by complex potentials for which one obtains:
\begin{eqnarray}
&&E=i\left(1-\frac{2M_1}{R_1}-\frac{2M_2}{R_2} \right),\nonumber\\
&&\Phi=-i\sqrt{2}e^{-2\phi}\left(\frac{Q_2}{R_2}+\frac{Q_1}{R_1}\right), \\
&&z=z_{\infty}-2ie^{-2\phi_{\infty}}\left(\frac{D_1}{R_1}+
\frac{D_2}{R_2} \right),\nonumber
\end{eqnarray}
where electric $Q_b$ and dilaton $D_b$ charges are introduced via
asymptotics
\begin{equation}
\phi \sim \phi_{\infty}+\frac{D}{R},\quad
v\sim \frac{\sqrt{2}Q e^{-\phi_{\infty}}}{R}.
\end{equation}
Masses are related to the seed ones as
\begin{equation}
M_1=\mu_1 \cosh^2\theta,\quad M_2=-\mu_2 \sinh^2\theta.
\end{equation}
Parameters $Q_b$ are not independent but satysfy
\begin{equation}
Q_1 Q_2=2M_1 M_2,
\end{equation}
and $D_b$ are related to other parameters as
\begin{equation}
D_b=-\frac{Q_b^2}{2M_b}.
\end{equation}
This relation is similar to that for the single dilaton black hole.
In the present case, however, total dilaton charge $D=D_1+D_2$ is
a free parameter for fixed $Q=Q_1+Q_2$ and $M_1+M_2$. Indeed,
$\theta$--parameter of the $S$--transformation (43) reads
\begin{equation}
\tanh 2\theta =\frac{\sqrt{2}Q}{M-D},
\end{equation}
and hence $D$ can be chosen as an independent quantity. Magnetic potential
is non--zero in the present solution, but there is no magnetic charge
(relevant combination $u-\kappa v$ does not possess Coulomb component).
If desired, magnetic charge can be generated using electric--magnetic
duality subgroup. Also, NUT--parameters can be either added to the seed
solution, or generated.

More general soliton solutions may be constructed by a direct application
of an inverse scattering transform method \cite{bz} to the matrix $M$.
Although we will not enter into details here, we give
the  parametrization of the coset in terms of real potentials which
follows from the representation (17) for matrix dilaton--axion
\begin{equation}
M=
f^{-1}\left(\begin{array}{crc}
M_1&M_3\\
M_3^T&M_2 \\
\end{array}\right),
\end{equation}
where
\[
M_1=
\left(\begin{array}{crc}
1&-v\\
-v&v^2-fe^{2\phi} \\
\end{array}\right),\quad
M_3=
\left(\begin{array}{crc}
-\chi&u\\
v\chi-wfe^{2\phi}&\kappa fe^{2\phi}-uv \\
\end{array}\right),
\]
\begin{equation}
M_2=
\left(\begin{array}{crc}
f^2+\chi^2-f\left(w^2e^{2\phi}+v^2e^{-2\phi}\right)&
f\left(\kappa we^{2\phi} -ve^{-2\phi}\right)-u\chi\\
f\left(\kappa we^{2\phi} -ve^{-2\phi}\right)-u\chi&
u^2-fe^{2\phi}\left(\kappa^2+e^{-4\phi}\right)\\
\end{array}\right).
\end{equation}
Under action of $Sp(4,R)$ the coset matrix  undergoes transformation (12)
with
\[
G_g=G_E^T=\left(\begin{array}{crc}
I&\lambda e_1\\
0&I\\
\end{array}\right),\quad
G_e=G_{H_1}^T(-\lambda)=\left(\begin{array}{crc}
I+\lambda e_3&0\\
0&I-\lambda e_4\\
\end{array}\right),
\]
\begin{equation}
G_s=\left(\begin{array}{crc}
e^{-\lambda} e_1+e_2&0\\
0&e^{\lambda} e_1+e_2\\
\end{array}\right),\quad
G_m=G_{H_2}^T=\left(\begin{array}{crc}
I&\lambda(e_3+e_4)\\
0&I\\
\end{array}\right),
\end{equation}
\[
G_{d_1}=G_{d_2}^T=\left(\begin{array}{crc}
I&-\lambda e_2\\
0&I\\
\end{array}\right),\quad
G_{d_3}=\left(\begin{array}{crc}
e^{-\lambda} e_2+e_1&0\\
0&e^{\lambda} e_2+e_1\\
\end{array}\right),
\]
where the following $2\times 2$ basis is used
\begin{equation}
e_1=\left(\begin{array}{crc}
1&0\\
0&0\\
\end{array}\right),\quad
e_2=\left(\begin{array}{crc}
0&0\\
0&1\\
\end{array}\right),\quad
e_3=\left(\begin{array}{crc}
0&1\\
0&0\\
\end{array}\right),\quad
e_4=
\left(\begin{array}{crc}
0&0\\
1&0\\
\end{array}\right).
\end{equation}

Under prime operation
\begin{equation}
M\rightarrow M^{'}=\epsilon M \epsilon,\quad \epsilon=
\left(\begin{array}{crc}
\varepsilon&0\\
0&\varepsilon\\
\end{array}\right),
\end{equation}
while under $Z$--inversion
\begin{equation}
M\rightarrow M^{''}=-M^{-1}.
\end{equation}

We conclude with the following remarks. Complex description
of the three--dimensional reduction of the EMDA system in terms of
Ernst--type potentials opens a convenient way for an application of an
inverse scattering transform method when theory is further reduced to two
dimensions. One may also effectively use discrete symmetries (prime
and double prime) in generating purposes. In fact we were able to derive
(although not the most general) two--soliton solution without invoking
the usual dressing technique.

It is worth noting that our Ernst--type potentials
do not reduce to the original EM Ernst potentials when $z=0$.
This is related to a different structure of the symmetry group:
symplectic group $Sp(4,R)$ instead of the pseudounitary one
$SU(2,1)$ in the EM case. The remarkable fact is that one can find
a representation for the main building block --- matrix dilaton--axion ---
linear in complex potentials and not containing extra variables
(contrary to the EM $SU(2,1)$ case \cite{ki}). In a sense, EMDA system
is even simpler than its EM prototype. One extra complex potential is
related to the gravitaional Ernst potential by a symmetry which
effectively reduces the complexity of the system. Finally we
have achieved matrix $SL(2,R)$ formulation which resembles
typical structure of the vacuum Einstein equations. In fact it was
already observed that black hole solutions to the EMDA system are more
similar to vacuum rather than to electrovacuum solutions. Now we can
understand deeper the nature of this property.
Coupling of gravity to the Maxwell field
changes an algebraic structure of three--dimensional
reduction of the theory. Surprisingly enough, addition of two scalar fields
restores the previous $SL(2,R)$ structure at a different level.

This work was supported in part by the Russian
Foundation for Fundamental Research Grant 93--02--16977,
and by the International Science Foundation and Russian Governement
Grant M79300.

\end{document}